\documentclass[aip,jap,reprint]{revtex4-1}      
\usepackage[utf8]{inputenc}
\usepackage{bm}
\usepackage[english]{babel}
\usepackage{graphics}

\newcommand\rara{\rangle\!\rangle}
\newcommand\lala{\langle\!\langle}

\begin{document}

\title{Non-Markovian magnetization dynamics for uniaxial nanomagnets}

\author{Pascal Thibaudeau}
\email{pascal.thibaudeau@cea.fr}
\affiliation{CEA DAM/Le Ripault, BP 16, F-37260, Monts, FRANCE}

\author{Julien Tranchida}
\email{julien.tranchida@cea.fr}
\affiliation{CEA DAM/Le Ripault, BP 16, F-37260, Monts, FRANCE}
\affiliation{CNRS-Laboratoire de Mathématiques et Physique Théorique (UMR 7350), Fédération de Recherche "Denis Poisson" (FR2964), Département de Physique, Université de Tours, Parc de Grandmont, F-37200, Tours, FRANCE}

\author{Stam Nicolis}
\email{stam.nicolis@lmpt.univ-tours.fr}
\affiliation{CNRS-Laboratoire de Mathématiques et Physique Théorique (UMR 7350), Fédération de Recherche "Denis Poisson" (FR2964), Département de Physique, Université de Tours, Parc de Grandmont, F-37200, Tours, FRANCE}

\date{\today}

\begin{abstract}
A stochastic approach for the description of the time evolution of the magnetization of nanomagnets is proposed, that interpolates between the Landau-Lifshitz-Gilbert and the Landau-Lifshitz-Bloch approximations, by varying the strength of the noise. Its finite autocorrelation time,  i.e. when it may be described as colored, rather than white, is, also,  taken into account and the consequences,  on the scale of the response of the magnetization are investigated. It is shown that the hierarchy for the moments of the magnetization can be  closed, by introducing a suitable truncation scheme, whose validity is tested by direct numerical solution of the moment equations and compared to the averages obtained from a numerical solution of the corresponding colored stochastic Langevin equation. This comparison is performed on magnetic systems subject to both an external uniform magnetic field and an internal one-site uniaxial anisotropy.
\end{abstract}

\maketitle

\section{Introduction}
Idealizations are  often employed, when  modeling real magnetization dynamics. Far from equilibrium, magnetic fluctuations, modeled by white noise are examples of such an idealization and the magnetization processes predicted in this way have been observed in ultrafast reversal magnetization experiments. However real fluctuations always have finite auto-correlation time and the corresponding spin system dynamics  is, rather, described by non-Markovian stochastic processes \cite{atxitia2009ultrafast}. Despite the practical relevance of  non-Markovian processes, however, these have received much less attention than Markovian processes, chiefly,  because of the simplicity of the theory and the familiarity of the corresponding mathematical tools \cite{Gardiner:1998vf}. The study of the magnetization dynamics of super-paramagnetic systems, however, is a suitable framework and provides motivation to describe magnetic fluctuations, induced by contact to a heat bath\cite{Garcia-Palacios:1998uq}, that are sensitive to non--Markovian effects.

The effective magnetization dynamics of a spin system and a noise is described by a stochastic differential equation of Langevin type, also known as the stochastic Landau-Lifshitz-Gilbert (sLLG) equation~:
\begin{equation}
(1+\lambda^2)\frac{\partial s_i}{\partial t}=\epsilon_{ijk}(\omega_j+ \Omega_j)s_k+\lambda (\omega_i s_j s_j-\omega_j s_j s_i). 
\label{LLG0}
\end{equation}
The Einstein summation convention is adopted, with Latin indices standing for vector components, $\epsilon_{ijk}$ is the anti-symmetric Levi-Civita pseudo-tensor and $\bm{s}$ is a magnetization 3-vector, whose norm is fixed and can be set to unity.  Here $\bm{\omega}$ is the precession frequency which derives from a spin Hamiltonian and is such connected to the spin vector. $\lambda$ is the damping coefficient. The stochastic properties of ${\bf \Omega}$ describe the  fluctuations of external and/or internal degrees of freedom, random perturbations and noise. Its correlation functions are assumed to satisfy the relations
\begin{equation}
\frac{\partial}{\partial t}\langle \Omega_{i}(t)\dots\Omega_{i_n}(t_n)\rangle=-\frac{1}{\tau}\langle\Omega_{i}(t)\dots\Omega_{i_n}(t_n)\rangle
\end{equation}
with $\langle\rangle$ and $t\ge t_1\ge \dots\ge t_n$ is the time-ordered statistical average product of stochastic variables \cite{Van-Kampen:1992it} and $\tau$ is a single characteristic constant time. With proper conditions Gaussian process, Poisson process and two-state (dichotomy) Markov process enjoy this ``time derivative property'' \cite{Shapiro:1978qa}. In general a detailed description of the stochastic processes in the extended phase space $({\bf s},{\bf \Omega})$ is not easy to establish~\cite{Miyazaki:1998ek,Thibaudeau:2012zr}.

Stochastic processes are completely described by an infinite set of multivariate probability distributions. In practice, not all multivariate distributions are not desirable but only a few, mainly the one-variable distribution $P({\bf s}, t)$ and the  two-variable distribution, $P({\bf s}_1,t_1 ; {\bf s}_2, t_2)$. The former is used  to determine the average  $\langle f({\bf s},t)\rangle$ of any functional $f$ of the process ${\bf s}(t)$, e.g. its statistical moments. The latter is used to determine two-time statistics, e.g., the correlation function $\langle s_i(t_1)s_j(t_2)\rangle$. In the white noise limit, a Fokker-Planck equation on $P({\bf s},t)$ has been established only for a thermally interacting spin system subject to a constant external field \cite{Brown-Jr:1963tp,Garcia-Palacios:1998uq}. Beyond the mean-field approximation, the explicit spin dependance of the effective field, including a magnetic anisotropy, on the probability distribution $P$, may be addressed.  

\section{Theory}
In contact with a thermostat, described by the stochastic vector, $\bm{\Omega}(t)$, the spin system precesses about the axis of magnetic effective field, defined by both a uniform external magnetic field with $\bm{\omega}^{(0)}=\gamma\mu_0\bm{H}$ and an anisotropic single-axis. This can be observed in super-paramagnetic nano-magnets which behave as effective, two-state, anisotropic spin systems \cite{Garcia-Palacios:1998uq}. The deterministic frequency vector is given by $\bm{\omega}=\bm{\omega}^{(0)}+\kappa \bm{n}\left(\bm{n}\cdot\bm{s}\right)$, where $\bm{n}$ is the anisotropy single-axis and $\kappa$ the anisotropy frequency.  The noise  vector, $\bm{\Omega}(t)$, is assumed to be described  by a Gaussian stochastic process, that is completely defined by its 1-point function $\langle\bm{\Omega}(t)\rangle$ and 2--point  function $\langle\Omega_i(t)\Omega_j(t')\rangle$. Therefore the only non--trivial information is contained in the these, that we assume to define an Ornstein-Uhlenbeck process~\cite{Uhlenbeck:1930kn}, which means that $\langle\Omega_i(t)\rangle=0$ and 
\begin{equation}
\langle\Omega_i(t)\Omega_j(t')\rangle=\frac{D}{\tau}\exp(-\frac{|t-t'|}{\tau})\delta_{ij}.
\end{equation}
It can be shown that  this process does converge to the centered white noise process, when $\tau\rightarrow 0$.

Moreover, it is assumed that the  thermostat can be described by a fluctuation-dissipation theorem that relates the strength of the noise, $D$, with the temperature $T$ by the relationship $D=\lambda k_bT/\hbar$, where $\lambda$ is the parameter that appeared in Eq.(\ref{LLG0}).

These microscopic degrees of freedom give rise to average quantities, that probe collective properties. Such quantities can be operationally defined by taking statistical averages over the noise of Eq.(\ref{LLG0}) at equal times. In the limit of low damping (i.e. $\lambda\ll 1$), the equation for the magnetization, $\left\langle s_i\right\rangle$, is~:
\begin{eqnarray}
\label{1pointfunction}
\frac{\partial \langle s_i\rangle}{\partial t}
&=&\epsilon_{ijk}(\langle \omega_js_k\rangle+\langle\Omega_js_k\rangle)\nonumber\\
&&+\lambda (\langle\omega_is_js_j\rangle-\langle\omega_js_js_i\rangle),
\end{eqnarray}
and implicates higher order correlation functions of noise and spin that have to be determined in turn. By applying the Shapiro-Loginov formulae of differentiation \cite{Shapiro:1978qa} to the components of the $\langle {\Omega}_i s_j \rangle$ at equal times, we obtain the equation
\begin{equation}
\frac{ \partial\langle{\Omega}_i s_j \rangle}{\partial t}= \left\langle {\Omega}_i \frac{\partial{s}_j}{\partial t} \right\rangle -\frac{1}{\tau}\left\langle {\Omega}_i s_j \right\rangle.
\label{Shapiro}
\end{equation}
Introducing the right-hand side of Eq.(\ref{LLG0}) into Eq.(\ref{Shapiro}), evolution equations for the mixed  moments $\langle {\Omega}_i s_j\rangle$ are obtained. Due to the non-linearity of the sLLG equation, an infinite hierarchy arises \cite{Nicolis:1998bd}: the equations for the second-order moments depend on third-order moments, and so on. In order to solve the system, closure relationships need to be found. So the equations for the moments $\langle s_i s_j\rangle$ are now required and are provided by $\partial\langle s_i s_j \rangle/\partial t = \langle s_i \dot{s}_j\rangle+\langle \dot{s}_i {s}_j \rangle$. Assuming that this identity holds, the time derivative of $s_i$ is replaced by the right-hand side of Eq.(\ref{LLG0}) inside the statistical average. This leads to an additional set of equations for the second-order moments of the microscopic spin degrees of freedom, that are closely related to those written by Garanin {\it et al} (see Eq.(6)\cite{Garanin:1990fk}) and Garcia-Palacios {\it et al.} (see Eq.(2.10)~\cite{Garcia-Palacios:1998uq}). As long as the stochastic variables follow individual Gaussian processes, their connected moments (labelled as $\lala . \rara$), beyond  the second moment are zero \cite{Van-Kampen:1992it}. The case of the mixed  connected moments of spin and noise is more subtle. If a linear coupling between the spin and the noise is assumed, the mixed cumulants, above  the second will, also, vanish. 
If  the Gaussian hypothesis is dropped, the general relations, at equal times, between the  desired moments are~:
\begin{eqnarray}
\langle\Omega_i s_j s_k\rangle&=&\langle\Omega_is_j\rangle\langle s_k\rangle+\langle\Omega_is_k\rangle\langle s_j\rangle+\lala\Omega_i s_j s_k\rara,\label{oss}\\
\langle\Omega_i s_j s_k s_l\rangle&=&\lala\Omega_i s_j s_k s_l\rara+\langle\Omega_is_j\rangle\langle s_k s_l\rangle+\langle\Omega_is_k\rangle\langle s_js_l\rangle\nonumber\\
&&+\langle\Omega_is_l\rangle\langle s_j s_k\rangle+\langle s_l\rangle\lala\Omega_i s_j s_k\rara\nonumber\\
&&+\langle s_k\rangle\lala\Omega_i s_j s_l\rara+\langle s_j\rangle\lala\Omega_i s_k s_l\rara,\label{osss}\\
\langle s_is_js_k\rangle&=&\langle s_is_j\rangle\langle s_k\rangle+\langle s_js_k\rangle\langle s_i\rangle+\langle s_is_k\rangle\langle s_j\rangle\nonumber\\
&&-2\langle s_i\rangle\langle s_j\rangle\langle s_k\rangle+\lala s_i s_j s_k\rara,\label{sss}\\
\langle s_is_js_ks_l\rangle&=&\langle s_is_j\rangle\langle s_ks_l\rangle+\langle s_is_k\rangle\langle s_js_l\rangle+\langle s_is_l\rangle\langle s_js_k\rangle\nonumber\\
&&-2\langle s_i\rangle\langle s_j\rangle\langle s_k\rangle\langle s_l\rangle+\langle s_k\rangle \lala s_i s_j s_l\rara\nonumber\\
&&+\langle s_j\rangle \lala s_i s_k s_l\rara+\langle s_l\rangle \lala s_i s_j s_k\rara\nonumber\\
&&+\langle s_i\rangle \lala s_j s_k s_l\rara+\lala s_i s_j s_k s_l\rara .\label{ssss}
\end{eqnarray}
These considerations lead to  the following system of equations~:
\begin{eqnarray}
\frac{\partial \langle s_i\rangle}{\partial t}
&=&\epsilon_{ijk}(\omega^{(0)}_j\langle s_k\rangle+\kappa n_jn_l\langle s_ls_k\rangle+\langle\Omega_js_k\rangle)\nonumber\\
&&+\lambda\left(\omega_i^{(0)}\langle s_js_j\rangle-\omega_j^{(0)}\langle s_js_i\rangle\right.\nonumber\\
&&+\big.\kappa n_l(n_i\langle s_ls_js_j\rangle-n_j\langle s_ls_js_i\rangle)\Big),\label{sys1}\\
\frac{\partial\langle\Omega_js_k\rangle}{\partial t}
&=&-\frac{1}{\tau}\langle\Omega_js_k\rangle+\epsilon_{klm}\left(
\langle\Omega_j\omega_ls_m\rangle+\langle\Omega_j\Omega_ls_m\rangle\right),\nonumber\\
&&+\lambda\left(\langle\Omega_j\omega_ks_ls_l\rangle-\langle\Omega_j\omega_ls_ls_k\rangle\right),\label{sys2}\\
\frac{\partial \langle s_is_j\rangle}{\partial t}
&=&\epsilon_{ikl}\left(\langle\omega_ks_ls_j\rangle+\langle\Omega_ks_ls_j\rangle\right)\nonumber\\
&&+\lambda\left(\langle\omega_is_ks_ks_j\rangle-\langle\omega_ks_ks_is_j\rangle\right)\nonumber\\
&&+(i\leftrightarrow j).\label{sys3}
\end{eqnarray}
These expressions are closed by considering first the independence of Gaussian processes, the form of the effective frequency vector and the expression of the autocorrelation function of the noise. One has explicitly~:
\begin{eqnarray}
\langle\Omega_j\omega_ls_m\rangle&=&\omega^{(0)}_l\langle\Omega_js_m\rangle\nonumber\\
&&+\kappa n_ln_u\left(\langle\Omega_js_u\rangle\langle s_m\rangle+\langle\Omega_js_m\rangle\langle s_u\rangle\right)\nonumber\\
\langle\Omega_j\Omega_ls_m\rangle&=&\frac{D}{\tau}\delta_{jl}\langle s_m\rangle\nonumber\\
\langle\Omega_j\omega_ks_ls_l\rangle&=&2\omega^{(0)}_k\langle\Omega_js_l\rangle\langle s_l\rangle\nonumber\\
&&+\kappa n_kn_u\left(\langle\Omega_js_u\rangle\langle s_ls_l\rangle+2\langle\Omega_js_l\rangle\langle s_us_l\rangle\right)\nonumber\\
\langle\Omega_j\omega_ls_ls_k\rangle&=&\omega^{(0)}_l\left(\langle\Omega_js_l\rangle\langle s_k\rangle+\langle\Omega_js_k\rangle\langle s_l\rangle\right)\nonumber\\
&&+\kappa n_ln_u\left( \langle\Omega_js_u\rangle\langle s_ls_k\rangle+\langle\Omega_js_l\rangle\langle s_us_k\rangle\right.\nonumber\\
&&+\left.\langle\Omega_js_k\rangle\langle s_us_l\rangle\right)\nonumber\\
\langle\omega_ks_ls_j\rangle&=&\omega^{(0)}_k\langle s_ls_j\rangle+\kappa n_kn_u\left(
\langle s_u\rangle\langle s_ls_j\rangle\right.\nonumber\\
&&\left.+\langle s_l\rangle\langle s_us_j\rangle+\langle s_j\rangle\langle s_us_l\rangle-2\langle s_u\rangle\langle s_l\rangle\langle s_j\rangle \right)\nonumber\\
\langle\Omega_ks_ls_j\rangle&=&\langle\Omega_ks_l\rangle\langle s_j\rangle+\langle\Omega_ks_j\rangle\langle s_l\rangle\nonumber\\
\langle\omega_is_ks_ks_j\rangle&=&\omega^{(0)}_i\left(\langle s_j\rangle\langle s_ks_k\rangle+2\langle s_k\rangle\langle s_ks_j\rangle\right.\nonumber\\
&&\left.-2\langle s_k\rangle\langle s_k\rangle\langle s_j\rangle
\right)+\kappa n_in_u\left(\langle s_ks_k\rangle\right.\nonumber\\
&&\left.\langle s_us_j\rangle+2\langle s_us_k\rangle\langle s_ks_j\rangle-2\langle s_u\rangle\langle s_k\rangle^2\langle s_j\rangle\right)\nonumber\\
%
\langle\omega_ks_ks_is_j\rangle&=&\omega^{(0)}_k\left(\langle s_j\rangle\langle s_ks_i\rangle+\langle s_k\rangle\langle s_is_j\rangle\right.\nonumber\\
&&\left.+\langle s_ks_j\rangle\langle s_i\rangle-2\langle s_k\rangle\langle s_i\rangle\langle s_j\rangle
\right)\nonumber\\
&&+\kappa n_kn_u\left(\langle s_us_k\rangle\langle s_is_j\rangle+\langle s_us_i\rangle\langle s_ks_j\rangle\right.\nonumber\\
&&\left.+\langle s_us_j\rangle\langle s_ks_i\rangle-2\langle s_u\rangle\langle s_k\rangle\langle s_i\rangle\langle s_j\rangle\right)\nonumber
\end{eqnarray}

Combined with Eqs.(\ref{sys1},\ref{sys2},\ref{sys3}), the derived system of equations is called the dynamical Landau-Lifshitz-Bloch model (dLLB). In the limit of $\tau\rightarrow 0$, Eq.(\ref{sys2}) gives a direct expression for the mixed moment of the noise and the magnetization as $\langle\Omega_i s_j\rangle=-D\epsilon_{ijk}\langle s_k\rangle$, which makes the $\langle\Omega_i s_j\rangle$ matrix antisymmetric. Once introduced into Eq.(\ref{sys1}), a longitudinal contribution  for the spin dynamics appears in which the amplitude of the noise is proportional to the square root of the temperature. For finite values of $\tau$, Eq.(\ref{sys1}) exhibits a "memory effect" which shows that the value of $\langle s_i(t)\rangle$ for the time $t$ depends on all the previous values of $\langle s_i(t')\rangle$ before the time $t$.

\section{Numerical experiments}
Numerical experiments were performed to test the consistency of the closure hypothesis. For ``low'' values of $\tau$, the dynamics is close to  Markovian and Eqs.(\ref{sys1},\ref{sys2},\ref{sys3}) are expected to match the results obtained from an averaged Markovian stochastic dynamics. In order to realize these comparisons, a reference model is constructed from Eq.(\ref{LLG0}) with a white noise random vector $\bm{\Omega}$. Multiple spin trajectories with the same initial conditions are generated and followed in time according to a previously detailed symplectic integration scheme \cite{Lewis:2003lq,Thibaudeau:2012zr,Beaujouan:2012zk}. A statistical average is taken over all the realizations. Eqs.(\ref{sys1},\ref{sys2},\ref{sys3}) are integrated with an eight-order Runge-Kutta integration scheme, with adaptive steps, constructed from the Jacobian of the whole system.

\begin{figure}[htb]
\centering
\resizebox{0.99\columnwidth}{!}{\includegraphics{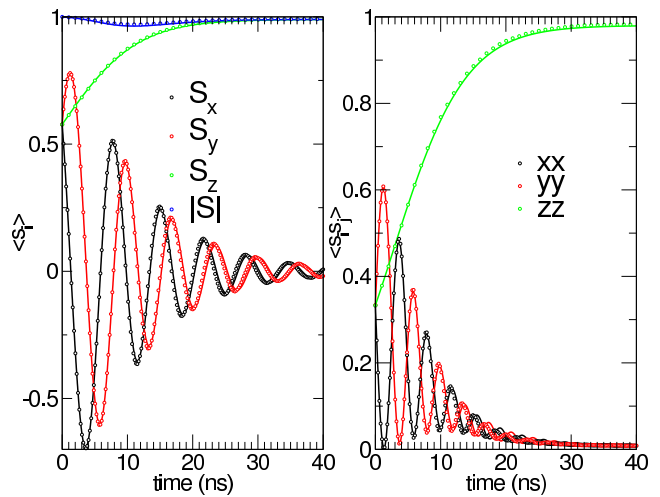}}
\caption{Dynamics of the averaged magnetization (left) and auto-correlation spin function (right) around a single-axis in the $z$-direction in the case of small correlation time of the noise. The dot lines plot the sLLG case ($10^3$ repetitions), solid lines stand for our dLLG model with the Gaussian hypothesis of independance. $D=1\times 10^{-3}{\rm{GHz}}; \lambda = 0.1; \tau=10^{-6}{\rm{ns}}; {\bf{\omega}}^{(0)}=(0,0,0); \kappa=1$GHz; ${\bf n}=(0,0,1)$ are taken with the initial conditions ${\bf s}(0)=(\frac{1}{\sqrt{3}},\frac{1}{\sqrt{3}},\frac{1}{\sqrt{3}})$, $\forall i,j$ $\langle s_i(0)s_j(0)\rangle=\frac{1}{3}$ and $\langle\Omega_i(0)s_j(0)\rangle=0$.}
\label{Fig1}
\end{figure}

\begin{figure}[htb]
\centering
\resizebox{0.99\columnwidth}{!}{\includegraphics{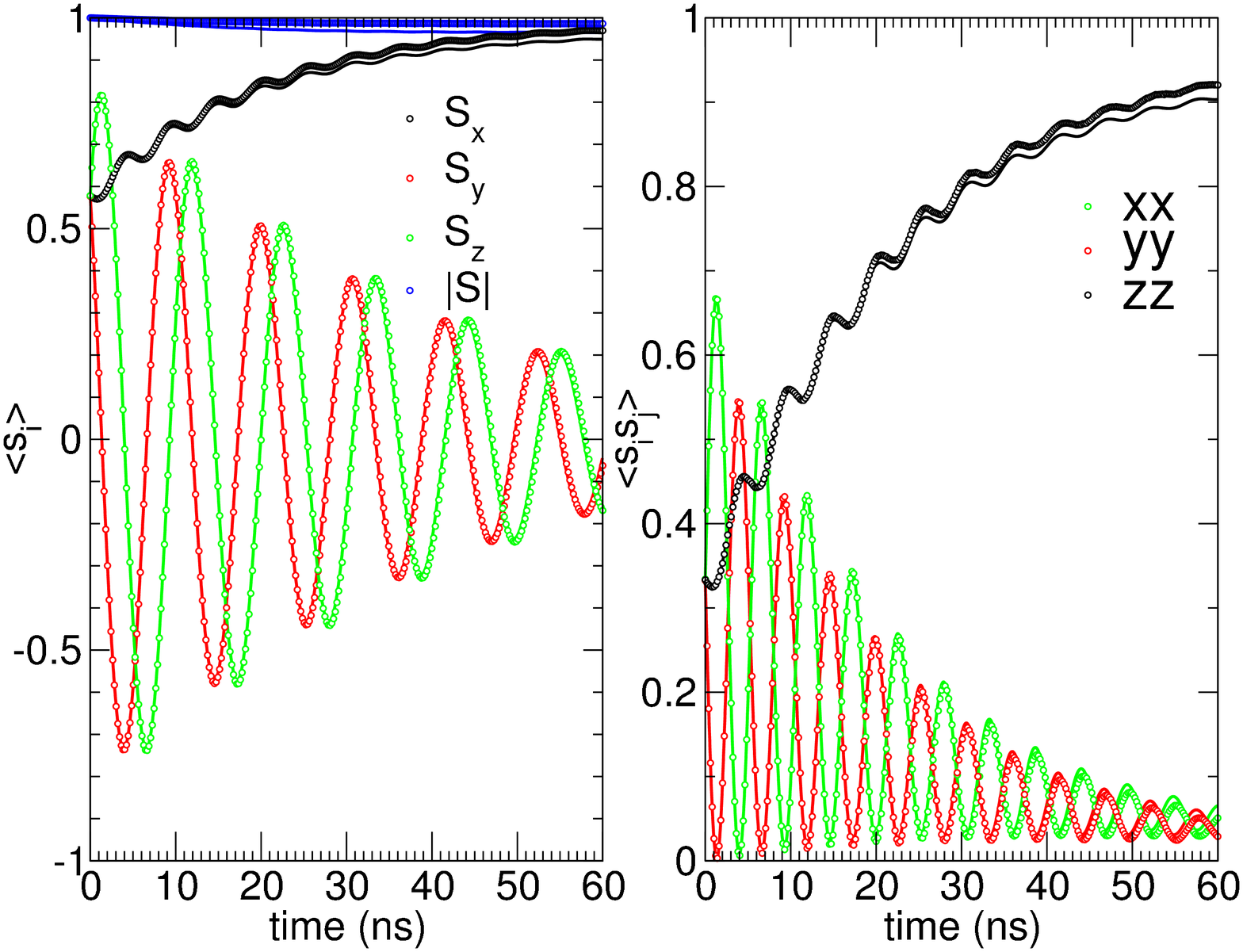}}
\caption{Stochastic average magnetization (left) and auto-correlation spin function (right) around a single-axis in the $z$-direction and a field in the $x$-direction. The dot lines are for the sLLG for $10^4$ repetitions and the solid lines stand for our dLLB model. $D=1\times 10^{-3}{\rm{GHz}}; \lambda = 0.05; {\bf{\omega}}^{(0)}=(\pi/5,0,0)$GHz ; $\kappa=1/10$GHz; ${\bf n}=(0,0,1)$; $\tau=10^{-1}$ ns are taken with the same initial conditions as previously.}
\label{Fig2}
\end{figure}

The figure (\ref{Fig1}-left) (resp. \ref{Fig1}-right) exhibits the average magnetization (resp. spin-spin autocorrelation function) from the stochastic LLG equation and our dLLB model for the same value of the amplitude $D$ in the white noise limit.
For low value of $D$ (i.e. in the normal phase), the average stochastic dynamics follows our dLLB model. A similar situation has been already investigated for an external constant field only \cite{Tranchida:hl}. When $\tau$ increases, the noise is no longer white and does not follow a Markov's process. A complete stochastic system of equations for both the spin and the noise is built by supplementing Eq.(\ref{LLG0}) with
\begin{equation}
\frac{d\Omega_i}{dt}=-\frac{1}{\tau}\left(\Omega_i-\xi_i(t)\right)
\label{stochnoise}
\end{equation}
where $\xi_i$ is a Gaussian centered stochastic process~\cite{Fox:1988tg}, but  with $\langle\xi_i(t)\rangle=0$ and $\langle\xi_i(t)\rangle\xi_j(t')\rangle=2D\delta_{ij}\delta(t-t')$. Formally, this system is integrated as $\Omega_i(t)\equiv e^{tL_{\Omega}}\Omega_i(0)=\int_0^te^{-(t-u)/\tau}\xi_i(u)du$ with $L_{\Omega}$ is the Liouville operator corresponding to Eq.(\ref{stochnoise}). Because the Liouville operators of the spin and the noise commute, an integration of the colored system is performed by the combination of the two operators as $e^{tL}(s_i(0),\Omega_i(0))=e^{tL_s}e^{tL_\Omega}(s_i(0),\Omega_i(0))=(s_i(t),\Omega_i(t))$. This is performed for larger values of $\tau$ for which the figure~(\ref{Fig2}) exhibits $\langle s_i(t)\rangle$ and the diagonal part of $\langle s_i(t)s_j(t)\rangle$.
It is observed that, even for ``large'' values of $\tau$, our dLLB model continues to track the dynamics of the colored stochastic LLG system. The influence of the higher than second-order cumulants on the dynamics of the average magnetization and spin autocorrelation function is mitigated  by a large damping factor $\lambda$ which prevents detailed descriptions of the magnetization dynamics. Moreover, because of the link between the amplitude $D$ and the damping factor $\lambda$, through a to-be-derived fluctuation-dissipation theorem in the special case of an explicit dependance of the spin in the effective field, the domain of validity of our dLLB model in temperature could be significantly controlled.
 
\begin{figure}[htb]
\centering
\resizebox{0.99\columnwidth}{!}{\includegraphics{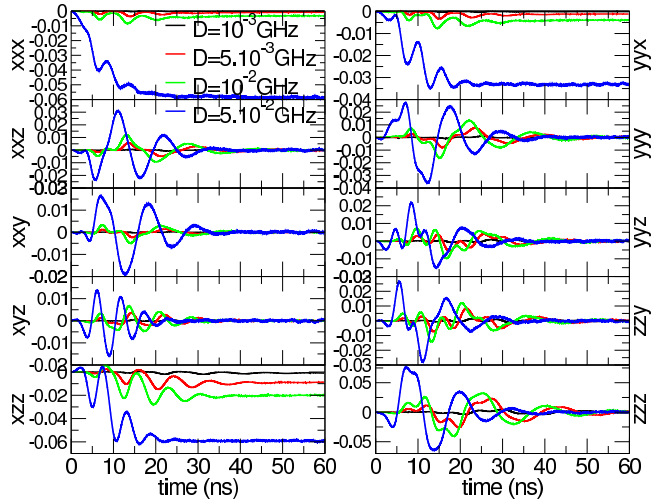}}
\caption{Stochastic average over for $10^5$ repetitions of the independent $\lala s_is_js_k\rara$ terms for various values of $D$. $\lambda = 0.1; {\bf{\omega}}^{(0)}=(\pi/5,0,0)$GHz ; $\kappa=0.5$GHz; ${\bf n}=(0,0,1)$; $\tau=10^{-3}$ns are taken with the same initial conditions as previously.}
\label{Fig3}
\end{figure}

When the intensity of the noise increases, the spin dynamics is no longer Gaussian. It is observed numerically, in the stochastic dynamics, by calculating the cumulant function $\lala s_i(t)s_j(t)s_k(t)\rara$. The ten independent spin cumulants are depicted, for several values of the amplitude $D$ in the figure (\ref{Fig3}).
As shown, for this particular situation, some terms go to zero in time, whereas some remain finite, depending on the underlying symmetries. At the expense of higher order expressions, the Eqs. (\ref{sys1},\ref{sys2},\ref{sys3}) can be supplemented by an expansion of $\frac{d\langle s_is_js_k\rangle}{dt}$ and integrated together. This represents an option where the deviation off the Gaussian character of the spin distribution is explicitly treated at the third-order. 

The third-order correlation functions for the dynamical variables ${\bf s}$ satisfying a stochastic Langevin-like equation are generally determined by first solving this equation in the presence of the noise and then averaging the product of these variables over the ensemble of the random noise. A path integral formalism can be established for deducing identities between  the correlation functions. This  strategy has been tested in various approximation schemes, including the quenched approximation (i.e. $D=0$). A shortcut might be to look for an alternate description of an effective system of equations in a closed time path formalism of Schwinger allowing to describe a general non-equilibrium field theory at finite temperature \cite{Kleinert:2006ye}. This is why a closure approximation was proposed by Martin, Siggia and Rose \cite{Martin:1973lr} which represents the classical limit of the closed time path formalism and will be the subject of further investigation.

JT acknowledges financial support through a joint doctoral fellowship ``Région Centre-CEA'' under grant number 00086667.

\end{document}